\documentclass{elsart}
\usepackage{graphics}
\usepackage{graphicx}
\usepackage{epsfig}
\usepackage{amssymb}

\begin{document}

\begin{frontmatter}

\title{Carbon doping of superconducting magnesium diboride}

\author{R.A. Ribeiro\corauthref{cor1}},
\corauth[cor1]{Corresponding author.} \ead{ribeiro@ameslab.gov}
\author{ S.L. Bud'ko},
\author{C. Petrovic},
\author{P.C. Canfield}
\address{Ames Laboratory and Department of Physics and Astronomy\\
Iowa State University, Ames, IA 50011 USA}

\begin{abstract}
We present details of synthesis optimization and physical
properties of nearly single phase carbon doped MgB$_{2}$ with a
nominal stoichiometry of Mg(B$_{0.8}$C$_{0.2}$)$_{2}$ synthesized
from magnesium and boron carbide (B$_{4}$C) as starting materials.
The superconducting transition temperature is $\approx$ 22 K
($\approx$ 17 K lower than in pure MgB$_{2}$). The temperature
dependence of the upper critical field is steeper than in pure
MgB$_{2}$ with $H_{c2}$(10K) $\approx$ 9 T. Temperature dependent
specific heat data taken in different applied magnetic fields
suggest that the two-gap nature of superconductivity is still
preserved for carbon doped MgB$_{2}$ even with such a heavily
suppressed transition temperature. In addition, the anisotropy
ratio of the upper critical field for $ T/T_c \approx \frac{2}{3}$
is $\gamma \approx$ 2. This value is distinct from 1 (isotropic)
and also distinct from 6 (the value found for pure MgB$_{2}$).
\end{abstract}

\end{frontmatter}

\section{Introduction}
Since the discovery of superconductivity in magnesium diboride at
elevated ($T_c \approx$ 40 K) temperature \cite{Jap1,Jap2},
considerable progress has been achieved in material synthesis as
well as in understanding of its physical properties
\cite{budkoPRL,DKF,AH,LANL,anisoPRB,anisoBRIF}. From these initial
days of research on superconducting MgB$_{2}$ many attempts were
made to tailor the physical properties of the material to suite
different needs as well as to explore the neighboring compounds in
search of even higher $T_{c}$ values.

A number of groups undertook synthesis and characterization of
(Mg$_{1-z}$T$_z$)B$_2$ or Mg(B$_{1-y}$M$_y$)$_2$ (T = transition
metal, Li, Be, Al; M = C, Si) materials. The agenda was
multi-fold: to look for changes in $T_c$, to perform tests of the
superconducting mechanisms in MgB$_2$, and to introduce additional
pinning centers that could lead to higher critical current
densities. Since many diborides crystallize in the same, hexagonal
AlB$_2$ type of structure as MgB$_{2}$, and these compounds have
been known and studied for decades \cite{borides} these
substitutions initially were viewed as feasible. In spite of
considerable efforts, substitutions in MgB$_2$ appeared to be
difficult and in many cases unsuccessful or, at best, ambiguous.

For magnesium site substitutions apparently only Al was shown to
enter the structure unambiguously \cite{Bob,Al1,Al2,Al3,Al4}
although in a limited concentration range. For boron site
substitutions a number of attempts with different elements were
made. Carbon substitution was reported in several publications
\cite{Paran,Ahn,Take,Mickel,Cheng,Bharati}. Most of these attempts
had elemental magnesium, boron and carbon as starting materials
and the synthesis was performed at different pressures and
temperatures. The results varied considerably depending on the
details of sample synthesis (for uniformity we refer to the
chemical formula written as Mg(B$_{1-x}$C$_x$)$_2$): carbon
solubility less then 1.25\% was reported in \cite{Paran}, a
two-step transition was observed in resistance measurements for
nominal $x = 0.1$ \cite{Ahn}, a solubility limit of approximately
$x = 0.35$ and shift of $T_c$ down to $\approx$ 34.8 K for $x =
0.03$ was reported in \cite{Take}, a $T_c$ value of 34 K (at 2\%
of full diamagnetic signal, $\Delta T_c =$ 3K) was measured by DC
magnetization at 20 Oe in Mg(B$_{0.8}$C$_{0.2}$)$_2$ (with no
mention of solubility limit) in \cite{Cheng}, and a solubility
limit of $x \approx 0.15$ and $T_c$($x = 0.15$) $\approx$ 30 K was
determined from magnetic measurements in \cite{Bharati}.

One of the difficulties associated with doping of MgB$_2$ may be
the fact that the MgB$_2$ structure is robust and an intimate,
atomic level, mixing of the dopant with the doped element before
or in the process of synthesis is required to achieve the
substitution. An interesting approach for carbon doping of MgB$_2$
was suggested by Mickelson \textit{et al.} \cite{Mickel}. The
starting materials for synthesis were elemental magnesium and
boron carbide (B$_4$C) powder. The result of their synthesis was
carbon doped MgB$_2$ as a majority phase and MgB$_2$C$_2$ as a
minority phase. The $T_c$ of the material was decreased by 7 K
(down to 32 K) as seen by magnetization and resistance
measurements. The resulting composition of the sample was
estimated to be Mg(B$_{0.9}$C$_{0.1}$)$_2$. This initial study
motivated us to attempt to optimize this synthetic route so as to
eliminate the impurity phases and to perform a thorough
investigation of the physical properties of the resulting
material.

\section{Experimental}

Samples of carbon doped MgB$_2$ for this study were synthesized in
the form of sintered pellets following the procedure used for pure
MgB$_2$ \cite{budkoPRL,DKF,ribeiro}. Magnesium lumps (99.9\%) and
B$_4$C powder (99\% - Alfa \AE sar) were sealed into tantalum
tubes, sealed in quartz, placed into a heated box furnace and then
(after the desired synthesis time) quenched to to room
temperature. For all the samples except one (MgB$_2$C$_{0.5}$),
the nominal stoichiometry was kept as Mg(B$_{0.8}$C$_{0.2}$)$_2$,
i. e.  Mg$_5$(B$_{4}$C)$_2$. Synthesis temperature and time were
varied systematically so as to optimize sample quality.

Powder X-ray diffraction (XRD) measurements were made at room
temperature using Cu $K_{\alpha}$ radiation in a Scintag
diffractometer. A Si standard was used for all runs. The Si lines
were removed from the X-ray diffraction data leading to the
apparent gaps in the powder X-ray data. The lattice parameters
were obtained by fitting the X-ray diffraction spectra using {\it
Rietica} software.

DC magnetization measurements were performed in Quantum Design
MPMS-5 and MPMS-7 SQUID magnetometers. Four-probe AC resistance
measurements were carried out in Quantum Design MPMS (with
external LR-400 and LR-700 resistance bridges) and PPMS-9 units.
Platinum wires were attached to the samples with Epotek H20E
silver epoxy. Heat capacity data was collected on small pressed
pellet samples using the PPMS-9 instrument in an applied field of
up to 9 T utilizing the relaxation technique.

\section{Synthesis Optimization}

\begin{figure}[htb]
\begin{center}
\includegraphics[angle=0,width=140mm]{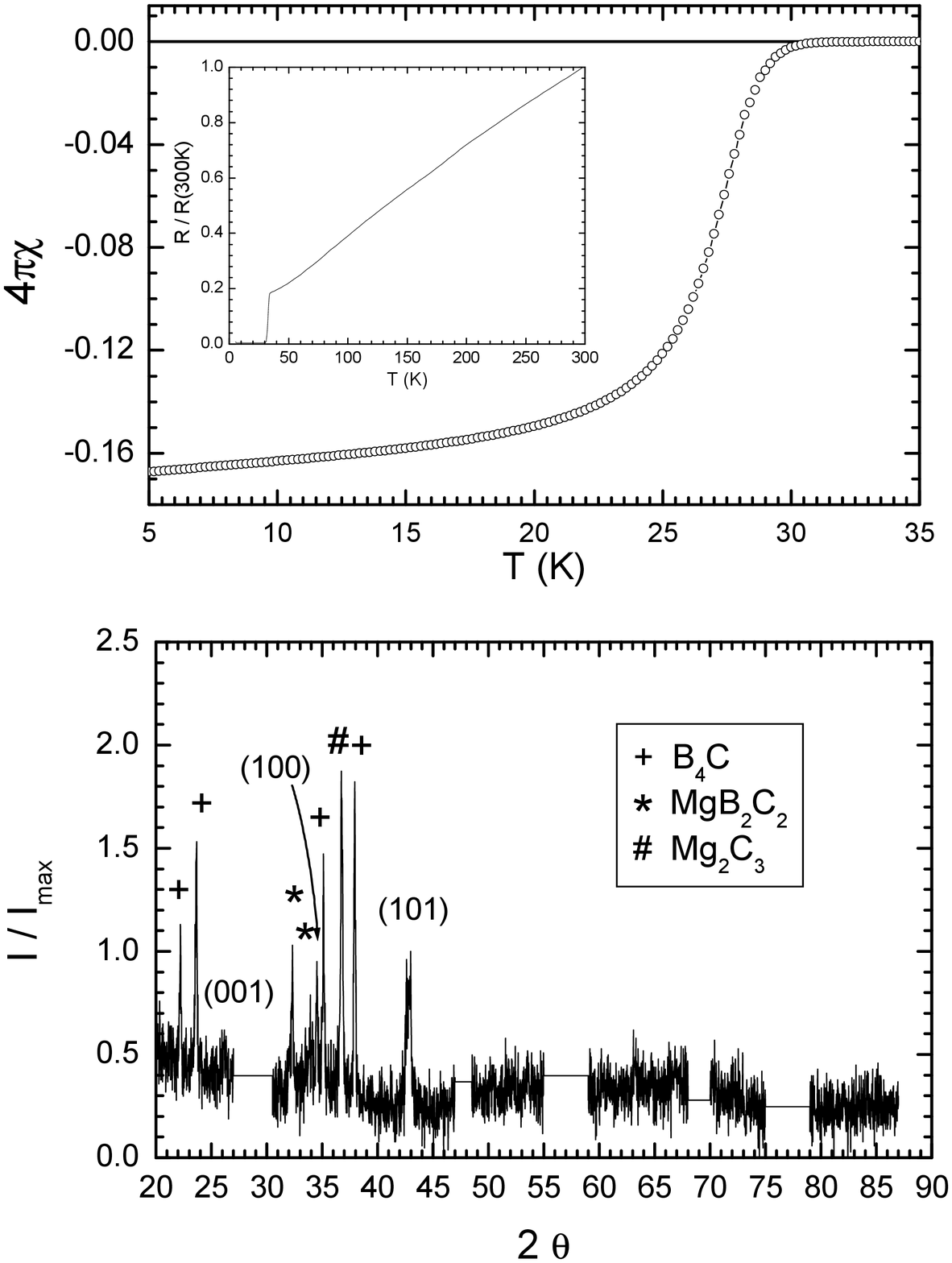}
\end{center}
\caption{{(a)temperature dependent magnetic susceptibility for
Mg(B$_{0.8}$C$_{0.2}$)$_2$ sample synthesized for 2 h at
600$^{\circ}$C and then for 2 h at 700$^{\circ}$C taken at $H$ =
50 Oe, ZFC - warming, insert - temperature dependent resistance
measured in zero applied field; (b) powder X-ray diffraction
pattern for the same sample. Numbers in parentheses - (hkl) for
Mg(B$_{1-x}$C$_{x}$)$_2$, symbols mark peaks from different
impurity phases.}}\label{Berkeley}
\end{figure}

Figure \ref{Berkeley}a presents low field magnetization data for a
nominal Mg(B$_{0.8}$C$_{0.2}$)$_2$ sample that was synthesized by
heating for two hours at $600^{\circ}$C and then for two more
hours at $700^{\circ}$C. This sample was made using the
temperature/time schedule outlined by Mickelson \textit{et al.}
\cite{Mickel} and serves as a point of comparison. $T_c$ of this
sample (defined via the onset of diamagnetism criterion) is
$\approx$ 30.5 K and the superconducting fraction is significantly
less than 100 \%. Temperature dependent resistance for this sample
is shown in the inset of Fig. \ref{Berkeley}a. The resistive
transition temperature is consistent with the one determined by
magnetic measurements. The residual resistance ratio $RRR$ =
$R$(300K)/$R_0 \approx$ 5.5 (with $R_0$ defined, in this case, as
normal state resistance just above the transition). Powder X-ray
diffraction (see Fig. \ref{Berkeley}b) confirmed that the
Mg(B$_{1-x}$C$_{x}$)$_2$ phase was formed, however three other
phases, Mg$_2$C$_3$, MgB$_2$C$_2$ and remnants of B$_4$C were also
detected. Although the $T_c$ of this sample is comparable with the
one reported by Mickelson \textit{et al.} \cite{Mickel}, it
appears to be poorly formed and clearly requires optimization.

\begin{figure}
\begin{center}
\includegraphics[angle=0,width=140mm]{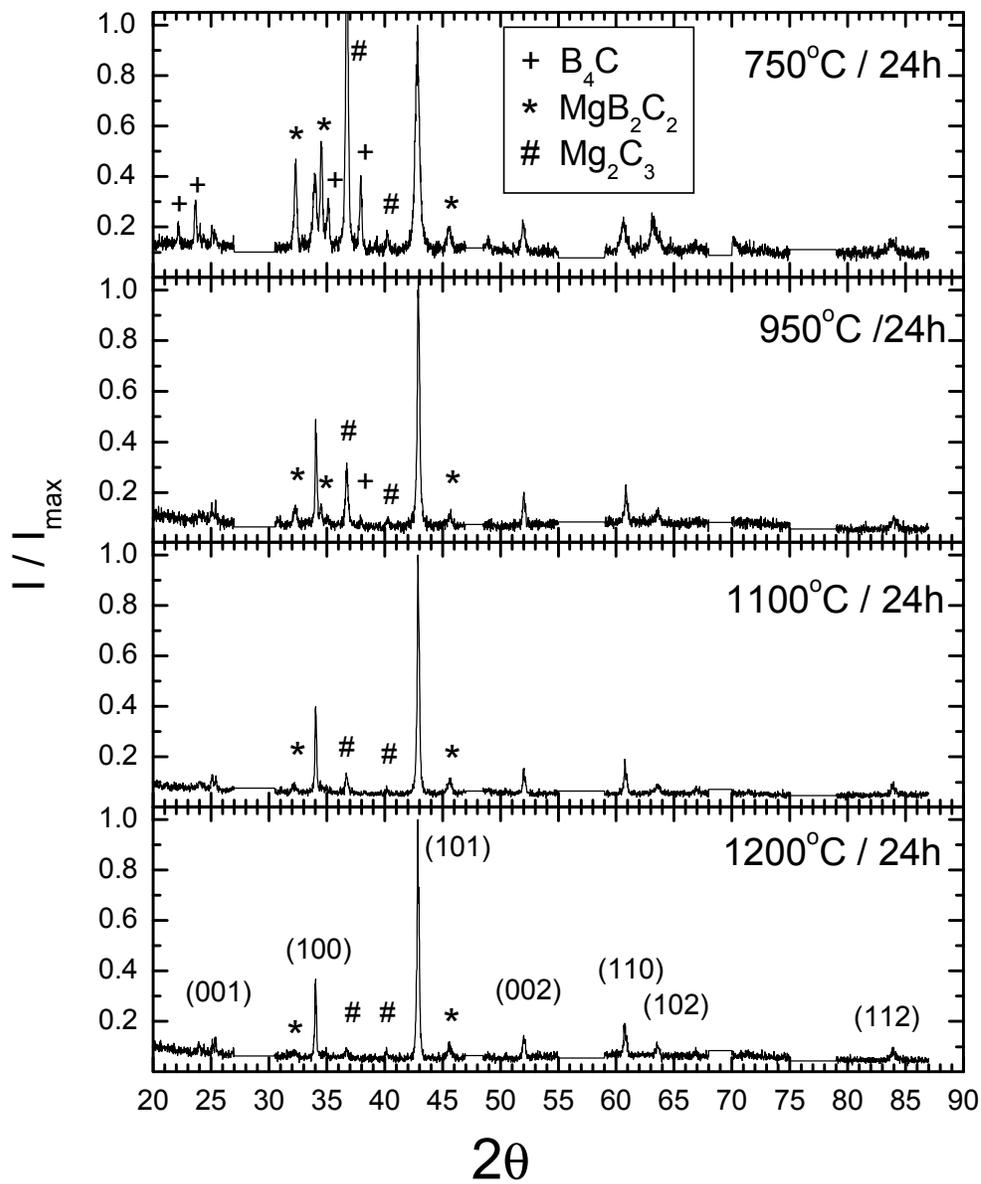}
\end{center}
\caption{Powder X-ray diffraction patterns for nominal
Mg(B$_{0.8}$C$_{0.2}$)$_2$ samples synthesized for 24 hours at
750$^{\circ}$C, 950$^{\circ}$C, 1100$^{\circ}$C, and
1200$^{\circ}$C. Numbers in parentheses in bottom panel - (hkl)
for Mg(B$_{1-x}$C$_{x}$)$_2$, symbols mark peaks from different
impurity phases: crosses - B$_4$C, asterisks - MgB$_2$C$_2$, pound
signs - Mg$_2$C$_3$.} \label{XRD24h}
\end{figure}

The presence of unreacted B$_4$C in the X-ray pattern indicates
that the reaction is probably not complete. Our next step in
optimization was to increase the reaction time to 24 hours and to
perform synthesis at number of different temperatures. Figure
\ref{XRD24h} presents powder X-ray diffraction spectra for nominal
Mg(B$_{0.8}$C$_{0.2}$)$_2$ samples synthesized for 24 hours at
four different temperatures: 750$^{\circ}$C, 950$^{\circ}$C,
1100$^{\circ}$C, and 1200$^{\circ}$C. Whereas the
750$^{\circ}$C/24h sample contains a considerable amount of
unreacted B$_4$C, traces of B$_4$C are much smaller for the
950$^{\circ}$C/24h sample and are not visible in the XRD patterns
of the 1100$^{\circ}$C/24h and 1200$^{\circ}$C/24h samples. In
addition the amount of the two other impurity phases (Mg$_2$C$_3$
and MgB$_2$C$_2$) clearly decrease with an increase in synthesis
temperature (see Fig. \ref{XRD24h}). The XRD patterns for
1100$^{\circ}$C/24h and 1200$^{\circ}$C/24h are very similar with
respect to the apparent quantities of the impurity phases and
present a significant improvement in purity in comparison to the
750$^{\circ}$C/24h and 950$^{\circ}$C/24h samples as well as to
the sample reported in ref. \cite{Mickel} and data presented in
Figure \ref{Berkeley}. The reactions carried out either at
1100$^{\circ}$C or 1200$^{\circ}$C appear to be approaching single
phase.

\begin{figure}
\begin{center}
\includegraphics[angle=0,width=140mm]{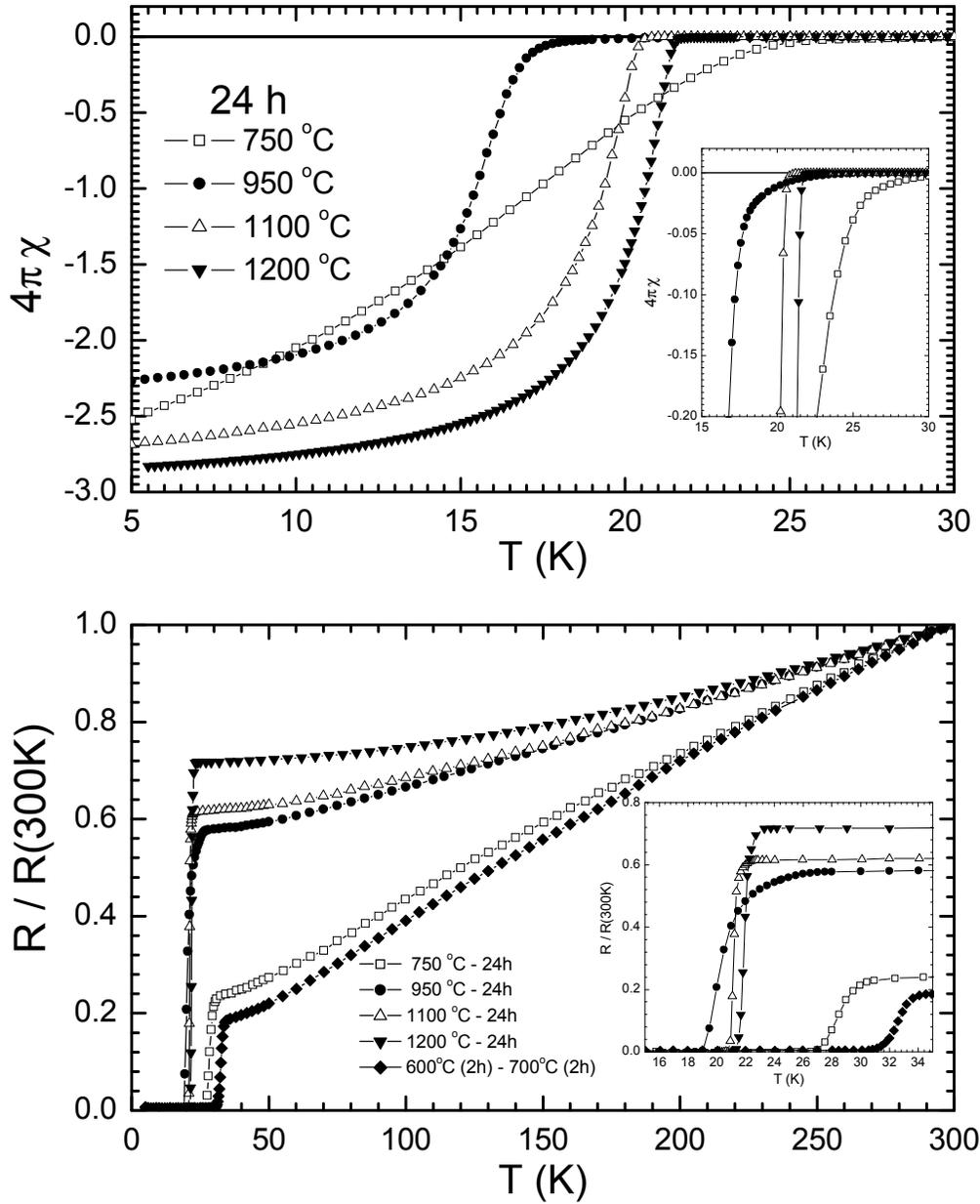}
\end{center}
\caption{(a)Temperature dependent magnetic susceptibility for
nominal Mg(B$_{0.8}$C$_{0.2}$)$_2$ samples synthesized for 24
hours at 750$^{\circ}$C, 950$^{\circ}$C, 1100$^{\circ}$C, and
1200$^{\circ}$C taken at $H$ = 50 Oe, ZFC - warming. Inset:
expanded temperature range near $T_c$. (b)Temperature dependent,
normalized resistance for the same set of samples plus reference
sample (see text). Inset: expanded temperature range near $T_c$.}
\label{MT24h}
\end{figure}

Low field DC magnetization and zero field resistance data taken
for the same set of samples also show an evolution of physical
properties with the synthesis temperature (Fig. \ref{MT24h}). The
superconducting transition for the 750$^{\circ}$C/24h sample has
an onset of the diamagnetic signal at 29 K but the transition, in
fact, is very broad that probably reveals a distribution of
transition temperatures within the sample probably due to chemical
inhomogeneities. The transitions as seen in the $M(T)$ data
sharpen with the increase of the reaction temperature. The onset
temperatures of the diamagnetic signal at the superconducting
transition seem to decrease with the increase of the reaction
temperature (see Fig. \ref{MT24h}a, inset), however the
temperatures at which the majority of the sample becomes
superconducting (50\% of the transition or maximum in $\partial
M/\partial T$ points) increase with the synthesis temperature for
950$^{\circ}$C/24h - 1200$^{\circ}$C/24h samples. Resistance data
(Fig. \ref{MT24h}b) manifest a similar trend e.g. the transition
width decreases with the increase of reaction temperature. The
transition temperatures defined using $R = 0$ criterion are 27,
19, 21, and 21.5 K for reaction temperatures of 750$^{\circ}$C,
950$^{\circ}$C, 1100$^{\circ}$C, and 1200$^{\circ}$C respectively.
$RRR$ decreases from 4.2 to 1.4 with the increase of the reaction
temperature.

At this point we attempt (with some hesitancy) to estimate the
room temperature resistivity or these samples. Very rough
evaluation results in the values: 0.3 m$\Omega$-cm, 0.4
m$\Omega$-cm, 2 m$\Omega$-cm and 2 m$\Omega$-cm for
750$^{\circ}$C/24h, 950$^{\circ}$C/24h, 1100$^{\circ}$C/24h, and
1200$^{\circ}$C/24h samples respectively. These numbers present
the {\it apparent} resistivity, with an understanding that (i) the
porosity of the samples was not taken into account, (ii) no
attempt was made to account for the possible contributions to the
measured resistivity value from grain boundaries and impurity
phases. Whereas the porosity of different samples prepared by the
route described above can be considered as similar and allow for
relative comparison of resistivities, we suggest that the possible
effects of grain boundaries and impurity phases cannot be reliably
accounted for within the available data on Mg(B$_{1-x}$C$_x$)$_2$
compounds and the minority phases encountered in them. Only gross,
order-of-magnitude, changes in apparent resistivity may, with some
reservations, be taken as reflecting the real evolution of
transport properties. Keeping this warning in mind we compare the
apparent room temperature resistivities of nearly single phase
nominal Mg(B$_{0.8}$C$_{0.2}$)$_2$ 1100$^{\circ}$C/24h and
1200$^{\circ}$C/24h compounds (2-3 m$\Omega$-cm) with those
measured on pellets of pure MgB$_2$ (0.02-0.03 m$\Omega$-cm
\cite{SHTSC}) synthesized by similar technique using isotopically
pure boron. In addition it should be mentioned that the apparent
resistivity for MgB$_2$ synthesized using only 90\% pure boron
\cite{ribeiro} was estimated to be 0.1 m$\Omega$-cm (these samples
have $RRR \approx$ 1.8, similar to carbon doped
1100$^{\circ}$C/24h and 1200$^{\circ}$C/24h compounds). In
addition, no literature reports on bulk, nominally pure MgB$_2$
prepared by different techniques report a room temperature
resistivity value above several tenths of a m$\Omega$-cm. Based on
these comparisons we conclude that our nearly single phase nominal
Mg(B$_{0.8}$C$_{0.2}$)$_2$ has a room temperature resistivity that
appears to be significantly higher than pure MgB$_2$. Any
conclusion beyond this fairly gross, qualitative statement runs
the risk of over interpreting these data.

\begin{figure}
\begin{center}
\includegraphics[angle=0,width=140mm]{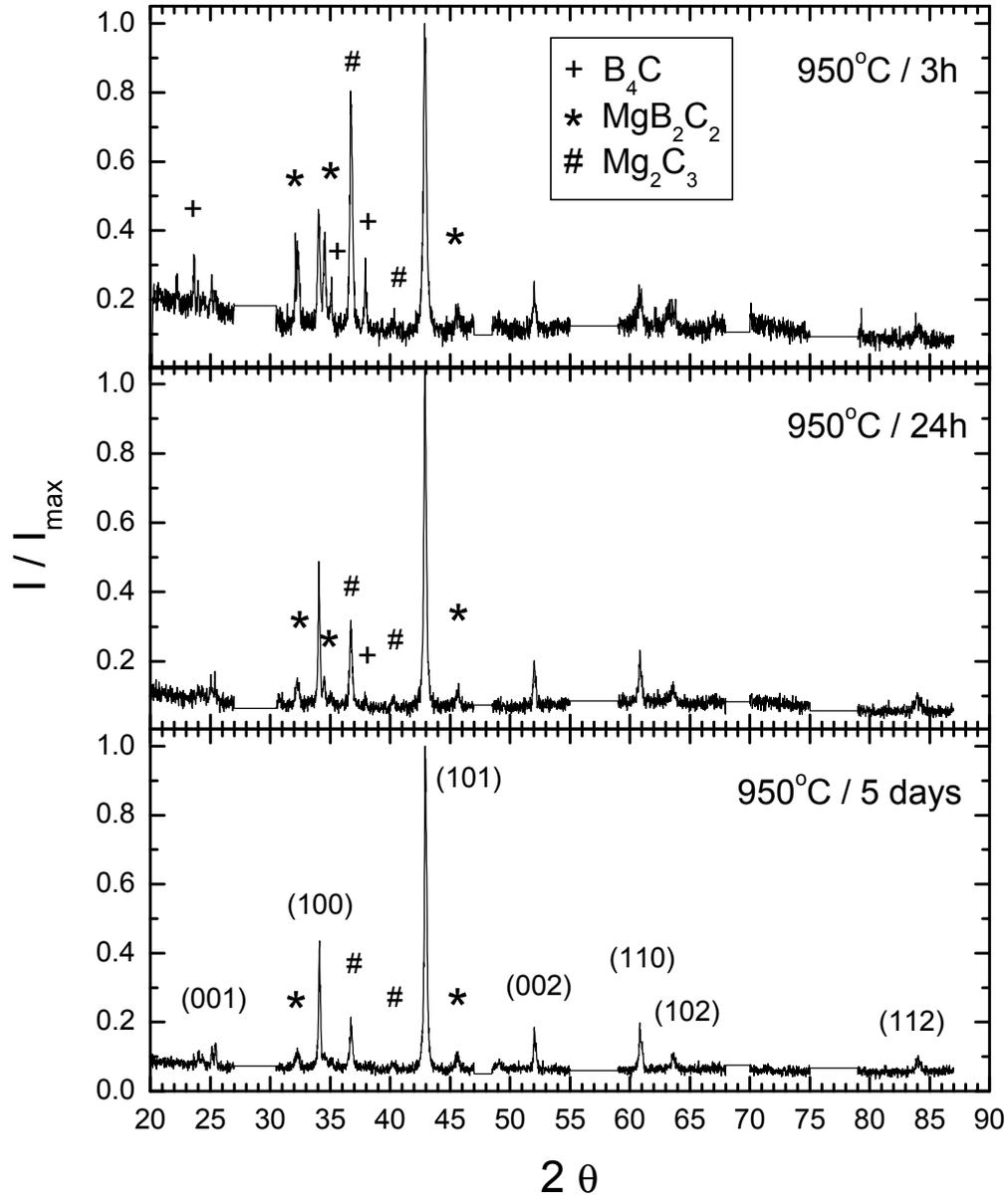}
\end{center}
\caption{Powder X-ray diffraction patterns for nominal
Mg(B$_{0.8}$C$_{0.2}$)$_2$ samples synthesized at 950$^{\circ}$C
for 3 h, 24 h, and 5 days. Numbers in parentheses in bottom panel
- (hkl) for Mg(B$_{1-x}$C$_{x}$)$_2$, symbols mark peaks from
different impurity phases.} \label{XRD950}
\end{figure}

\begin{figure}
\begin{center}
\includegraphics[angle=0,width=140mm]{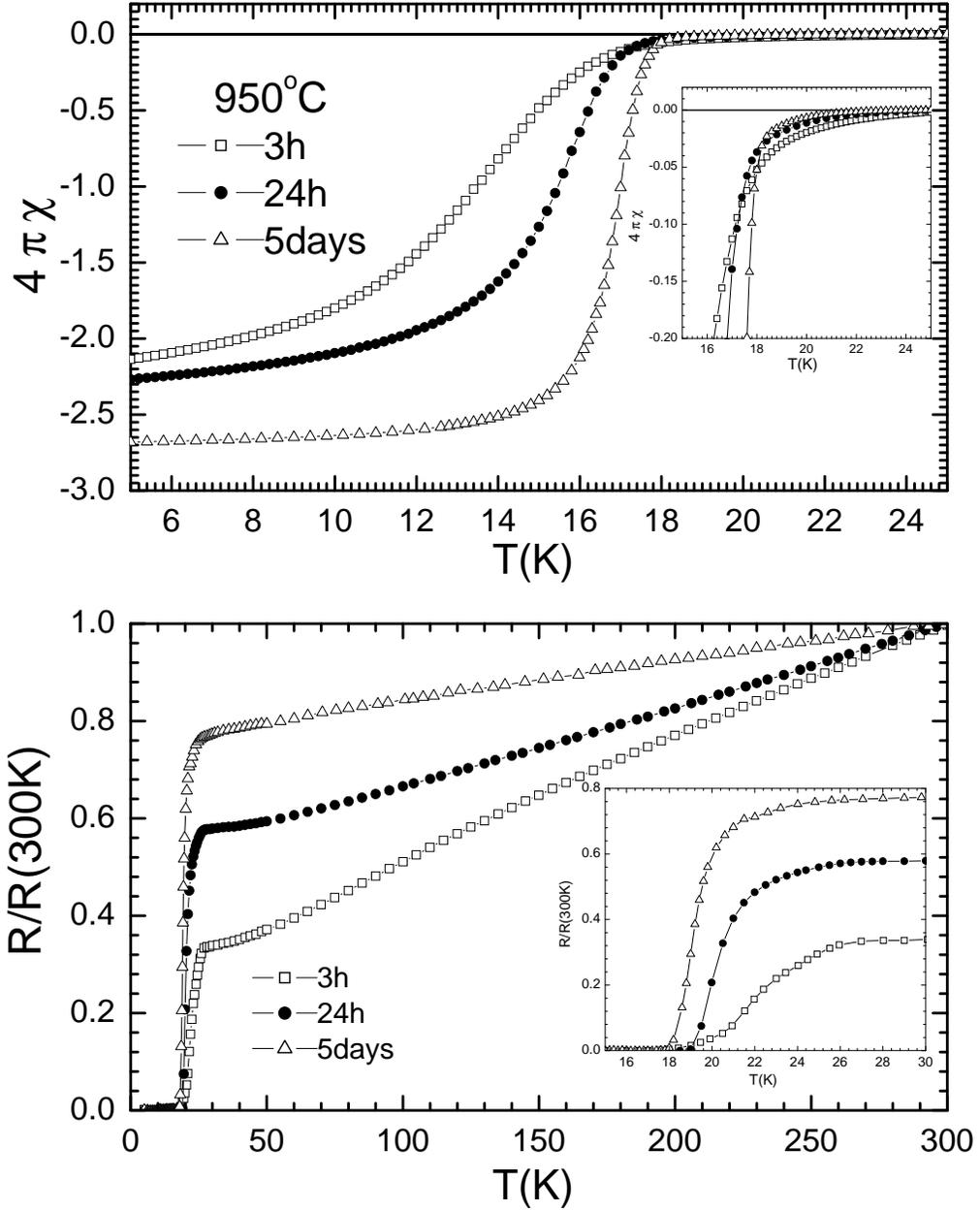}
\end{center}
\caption{(a)Temperature dependent magnetic susceptibility for
nominal Mg(B$_{0.8}$C$_{0.2}$)$_2$ samples synthesized at
950$^{\circ}$C for 3 h, 24 h, and 5 days. (b)Temperature dependent
normalized resistance for the same set of samples. Inset: expanded
temperature range near $T_c$.} \label{MT950}
\end{figure}

Since in a number of applications synthesis at lower temperatures
can be beneficial and because the details of the reaction between
elemental Mg with B$_4$C are not known, we have checked to see if
by performing this reaction at lower temperatures but longer time
we can obtain a material of similar or superior quality to the
1100$^{\circ}$C/24h - 1200$^{\circ}$C/24h samples. X-ray
diffraction patterns taken on samples reacted at 950$^{\circ}$C
for 3 hours, 24 hours and 5 days are shown in Fig. \ref{XRD950}.
For this reaction temperature the samples tend to improve with
increasing reaction time, the intensities of the peaks
corresponding to impurity phases monotonically decrease as
reaction time increases from 3 h to 24 h to 5 days. The traces of
unreacted B$_4$C are not seen for 950$^{\circ}$C/5 days sample but
Mg$_2$C$_3$ and MgB$_2$C$_2$ are still clearly detectable. The
quality of this sample (as inferred from XRD data) is approaching
that of the sample synthesized at 1100$^{\circ}$C or
1200$^{\circ}$C for 24 hours (compare Figs. \ref{XRD24h} and
\ref{XRD950}) but is still apparently inferior. Similarly, the low
field magnetization (see Fig. \ref{MT950}a) shows a gradual
decrease of the width of the transition, indicating more
homogenous samples, with the increase of the reaction time.
Temperature dependent resistance data (Fig. \ref{MT950}b) follow
the same trend: transition width decreases with increases in the
reaction time. The resistive transition temperatures (from $R$ =
0) are 19, 19, and 18 K for the 3 h, 24 h, and 5 days reaction
times. The $RRR$ values decrease from 3 to 1.3 with the increase
of the reaction time. On the other hand the transitions seen for
all reaction times, even the 950$^{\circ}$C/5 days sample, are all
significantly broader than those seen for the 1100$^{\circ}$C/24 h
and 1200$^{\circ}$C/24h samples.

\begin{figure}
\begin{center}
\includegraphics[angle=0,width=140mm]{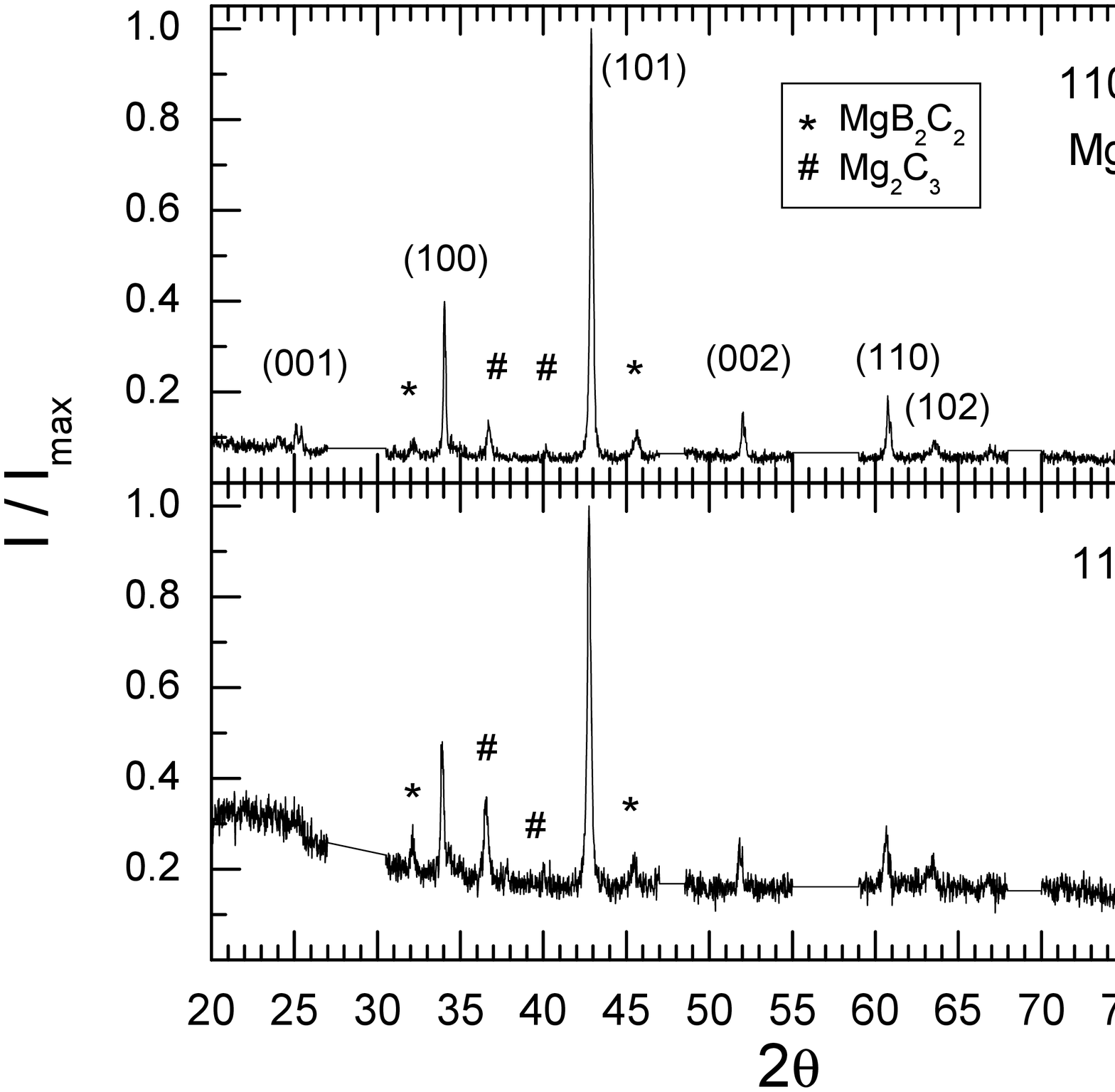}
\end{center}
\caption{Powder X-ray diffraction patterns for nominal
Mg(B$_{0.8}$C$_{0.2}$)$_2$ and MgB$_2$C$_{0.5}$ samples
synthesized at 1100$^{\circ}$C for 24 h. Numbers in parentheses
upper panel - (hkl) for Mg(B$_{1-x}$C$_{x}$)$_2$, symbols mark
peaks from different impurity phases: crosses - B$_4$C, asterisks
- MgB$_2$C$_2$, pound signs - Mg$_2$C$_3$.} \label{XRDsto}
\end{figure}

\begin{figure}
\begin{center}
\includegraphics[angle=0,width=140mm]{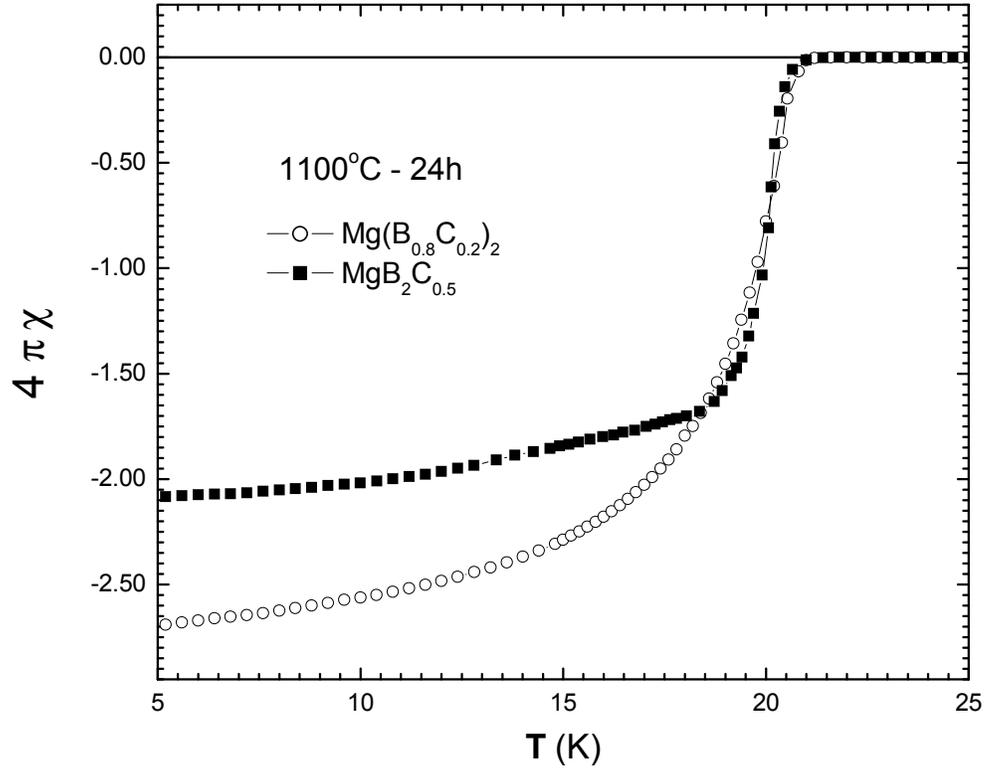}
\end{center}
\caption{Temperature dependent magnetic susceptibility ($H$ = 50
Oe, ZFC - warming) for nominal Mg(B$_{0.8}$C$_{0.2}$)$_2$ and
MgB$_2$C$_{0.5}$ samples synthesized at 1100$^{\circ}$C for 24 h.}
\label{MTsto}
\end{figure}

Based on the analysis of the two aforementioned data sets, the
1100$^{\circ}$C/24h - 1200$^{\circ}$C/24h reactions appear to
optimize the sample quality. Having this in mind we wanted to
address one final synthesis concern: would samples with nominal
stoichiometry of Mg(B$_{0.8}$C$_{0.2}$)$_2$ or MgB$_2$C$_{0.5}$ be
cleaner, i. e. do we get better samples with a Mg:B ratio of 1:2
or a Mg:(B + C) ratio of 1:2? Both types of samples we synthesized
following the 1100$^{\circ}$C/24h schedule. Powder XRD spectra
(Fig. \ref{XRDsto}) attest to the fact that the
Mg(B$_{0.8}$C$_{0.2}$)$_2$ sample is cleaner. Low field
temperature dependent magnetization (Fig. \ref{MTsto}) show that
the transition temperatures of these two samples are virtually
identical with a possibly slightly higher superconducting fraction
found for the Mg(B$_{0.8}$C$_{0.2}$)$_2$ sample. Apparently the
change to a MgB$_2$C$_{0.5}$ initial stoichiometry does not
improve the purity of the phase or sharpness of the
superconducting transition but does promote unwanted second
phases.

To summarize, the analysis of the three sets of samples discussed
above (having reaction temperature, reaction time and the Mg :
B$_4$C ratio as variables) we find that within the limitations of
our synthesis route the Mg(B$_{0.8}$C$_{0.2}$)$_2$ samples reacted
at 1100$^{\circ}$C/24h - 1200$^{\circ}$C/24h are the closest to
being single phase samples and have sharper superconducting
transition as seen in the temperature dependent resistance and low
field magnetization data. The transition for these samples has
shifted down by $\approx$ 17 K ($T_c \approx$ 22 K) with respect
to pure MgB$_2$. The $a$-lattice parameter is approximately 1.2\%
smaller than for pure MgB$_2$ whereas the $c$-lattice parameter
remains practically unchanged, a trend consistent with previous
data \cite{Take,Mickel,Cheng,Bharati}. The phase purity of these
samples is better than that of the sample reported by Mickelson
\textit{et al.} \cite{Mickel} and clearly much better than the
sample described in Figure \ref{Berkeley}. From the resistance
data for the two sets of samples (with reaction temperature or
reaction time being varied) it appears that the more phase-pure
and homogenous samples have lower residual resistance ratios. This
observation is opposite to what follows from the Mattheissen's
rule and is consistent with significant extrinsic contribution to
the normal state resistance (and $RRR$) from other phases, grain
boundaries, etc. that render attempts to use even semiquantitative
arguments based on normal state transport properties, e. g. the
Testardi correlation \cite{Bharati},  for such samples ambiguous.
On the other hand, it is fairly clear that the intrinsic
resistivity of the Mg(B$_{0.8}$C$_{0.2}$)$_2$ samples is
significantly higher than that of our pure MgB$_2$ samples
\cite{DKF,AH,ribeiro}.

A serious shortcoming of carbon doping through this (Mg + B$_4$C)
reaction route is that determination of the carbon content in the
final sample appears to be a difficult and non-trivial task. In
this report we will not attempt to address this issue for our
samples and will leave this problem for future work. It should be
noted that reliable and accurate quantitative determination of the
carbon content ($x$) and/or {\it consistent} relation between
three parameters: $T_c$, $x$ and $a$-lattice parameter (since $c$
is practically constant in all reports) cannot be found in the
available literature \cite{Paran,Ahn,Take,Mickel,Cheng,Bharati}.
As an aside note we would like to mention that we attempted to
dope MgB$_2$ with silicon and phosphorus through (Mg + B$_6$Si)
and (Mg + B$_{13}$P$_2$) synthesis routes. These attempts were
apparently not successful and resulted in multiphased compounds
with no indication of doping into B site.

\section{Physical Properties}

\begin{figure}
\begin{center}
\includegraphics[angle=0,width=140mm]{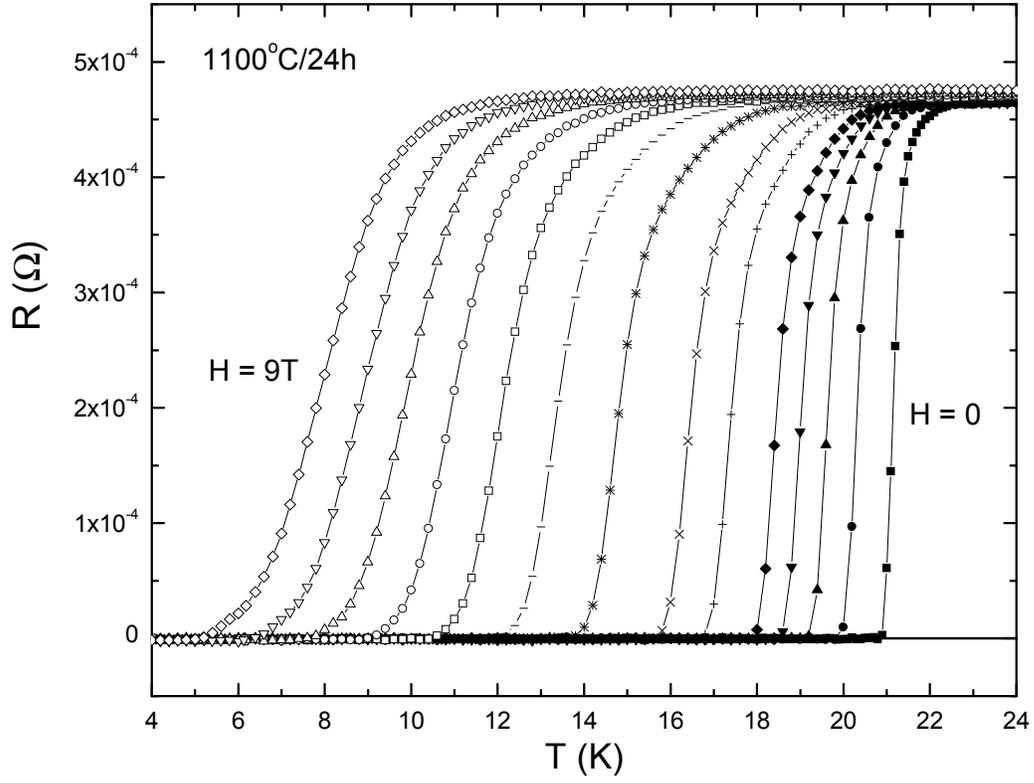}
\end{center}
\caption{Temperature dependent resistance for 1100$^{\circ}$C/24h
sample in applied magnetic field (fields from right to left: 0,
0.25, 0.5, 0.75, 1, 1.5, 2, 3, 4, 5, 6, 7, 8, 9 T). Region near
the transition shown.} \label{RT}
\end{figure}

\begin{figure}
\begin{center}
\includegraphics[angle=0,width=140mm]{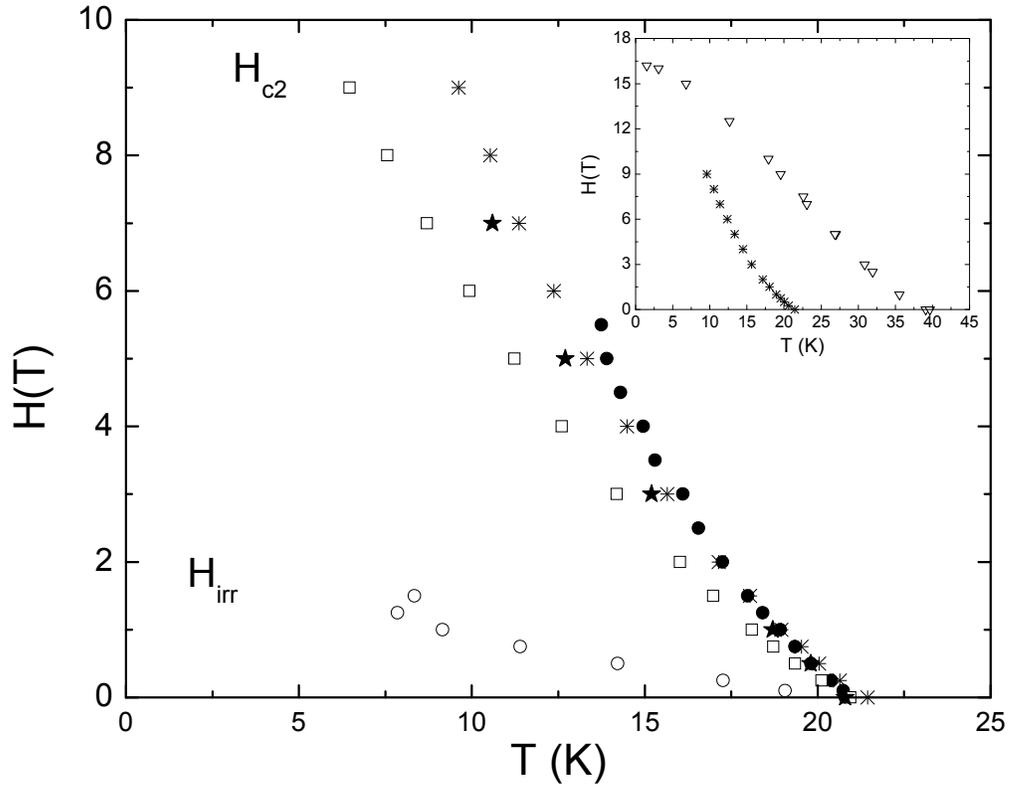}
\end{center}
\caption{Upper critical field and irreversibility line for
Mg(B$_{0.8}$C$_{0.2}$)$_2$ --- 1100$^{\circ}$C/24h sample. For
$H_{c2}$: filled circles - from magnetization, asterisks - from
onset criterion of magnetotransport data, squares - from $R$ = 0
criterion of magnetotransport data and stars - from heat capacity.
Inset: comparison of $H_{c2}$ for pure MgB$_2$ \cite{LANL} (open
triangles) and carbon doped sample.} \label{Hc2}
\end{figure}

\begin{figure}
\begin{center}
\includegraphics[angle=0,width=140mm]{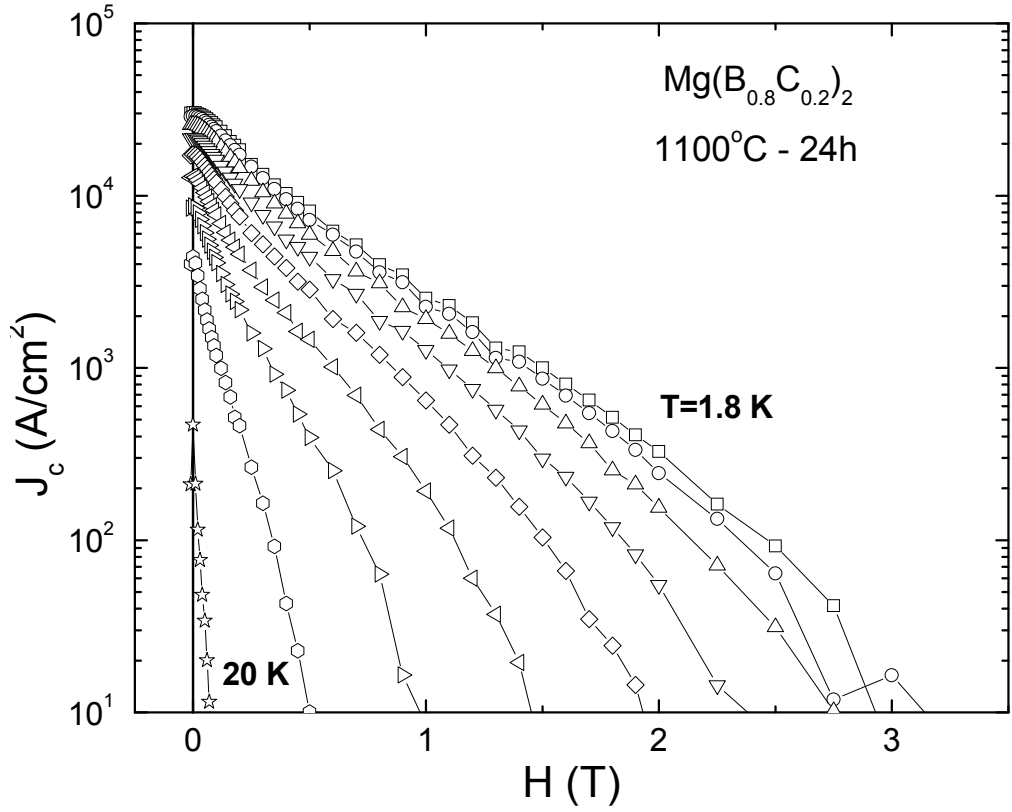}
\end{center}
\caption{Critical current density for Mg(B$_{0.8}$C$_{0.2}$)$_2$
--- 1100$^{\circ}$C/24h sample as inferred from magnetization
loops. Temperatures, from right to left: 1.8, 3, 5, 7.5, 10, 12.5,
15, 17.5, $20 K$.} \label{Jc}
\end{figure}

Once the best available synthesis route was established, the
nearly single phase 1100$^{\circ}$C/24h sample was chosen for more
detailed measurements of physical properties. Fig. \ref{RT}
presents the temperature dependent resistance measurements taken
in different applied fields up to 9 T (an enlarged region near the
superconducting transition in 0 - 9 T field range is shown).
Unlike the case of pure MgB$_2$ samples \cite{DKF,AH,LANL}
practically no magnetoresistance in the normal state was observed.
This is consistent with the very low residual resistance ratio,
$RRR \approx$ 1.6 for this sample and high estimated $\rho_0$. the
magnetotransport data (Fig. \ref{RT}) together with field and
temperature dependent magnetization data (not shown here) were
used to determine the upper critical field for this sample (Fig.
\ref{Hc2}). The irreversibility line ($H_{irr}$) shown in this
figure was determined from $M(H)$ loops taken at different
temperatures. The irreversibility field for this sample is quite
low: it extrapolates to $H_{irr}(0) \approx$ 2 T, which probably
points to the fact that the carbon substitutions in the sample
prepared from the Mg and B$_4$C mixture at 1100$^{\circ}$C/24h
conditions do not significantly increase the pinning. On the other
hand, the $H_{c2}(T)$ slope for this sample is considerably
steeper than in pure MgB$_2$ (see Fig. \ref{Hc2}, inset), so
although $T_c$ of the carbon doped sample is approximately half of
that for pure MgB$_2$, the extrapolation $H_{c2}(T \rightarrow 0)$
will give the value close to 16 T, similar to that of high purity
MgB$_2$ \cite{LANL}. Fig. \ref{Jc} presents the critical current
density for the 1100$^{\circ}$C/24h sample as determined from
magnetization loops using the Bean model \cite{Bean}. The critical
current densities are quite low, $J_c$(1.8K, $H$=0) $\approx$ 30
kA/cm$^2$, which is consistent with low pinning and the low lying
irreversibility line (Fig. \ref{Hc2}).

\begin{figure}
\begin{center}
\includegraphics[angle=0,width=140mm]{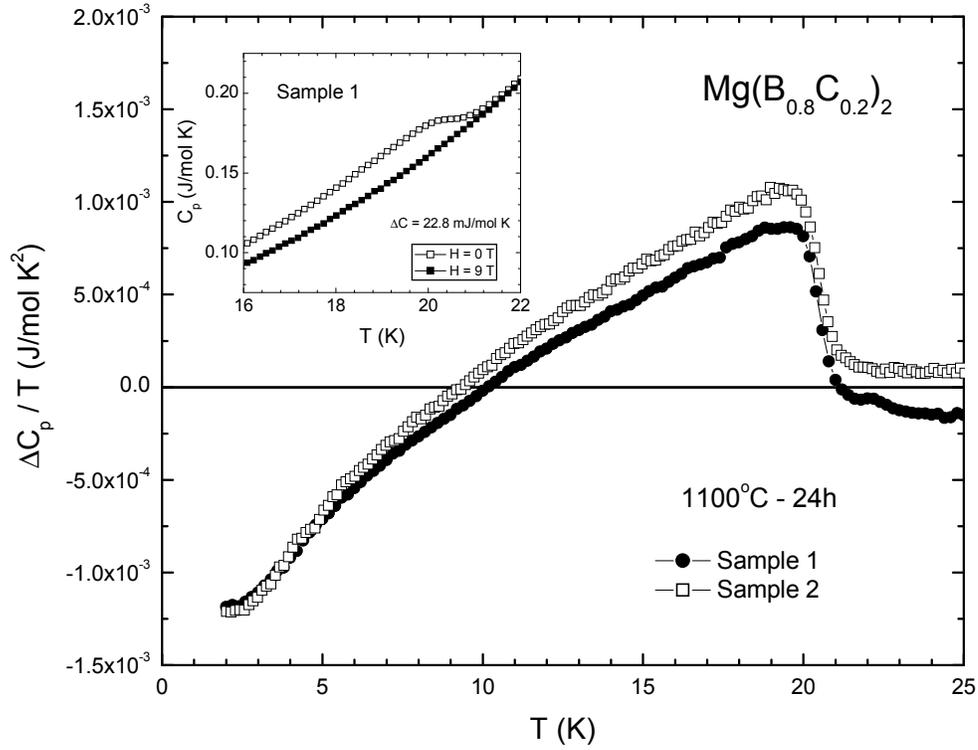}
\end{center}
\caption{Heat capacity difference $\Delta C_p/T =
(C_p(H=0)-C_p(9T))/T$ as a function of temperature for two
different Mg(B$_{0.8}$C$_{0.2}$)$_2$ --- 1100$^{\circ}$C/24h
samples. Inset: $C_p(T)$ data for $H$ = 0 and $H$ = 9 T in the
region near $T_c$.} \label{HC_1}
\end{figure}

The heat capacity of the 1100$^{\circ}$C/24h nominal
Mg(B$_{0.8}$C$_{0.2}$)$_2$ sample was measured on two different
pressed pellets, from two separate batches in zero and 9 T applied
field. The specific heat jump at the superconducting transition is
clearly seen (Fig. \ref{HC_1}, inset) and the value of the jump is
estimated as $\Delta C \approx$ 23 mJ/mol K. From 9 T measurements
the electronic term in specific heat is extrapolated to $\gamma
\approx$ 1.9 mJ/mol K$^2$ and the Debye temperature is estimated
as $\Theta_D \approx$ 685 K. Both $\gamma$ and $\Theta_D$ values
are lower than those accepted in the literature for pure MgB$_2$
samples \cite{budkoPRL,HC1,HC2,HC3} and have the same trend as
seen experimentally in \cite{Bharati} and theoretically in
\cite{med}. At a gross level, the significant decrease in $T_c$ is
consistent with lower $\gamma$ and $\Theta_D$ values for the
carbon doped sample. The heat capacity difference $\Delta C_p/T =
(C_p(H=0)-C_p(9T))/T$ as a function of temperature for the two
different samples is plotted in Fig. \ref{HC_1}. Both
qualitatively and quantitatively the two sets of data are similar
but it is worth noting that there is some sample-to-sample and
measurement-to-measurement variation.

\begin{figure}
\begin{center}
\includegraphics[angle=0,width=140mm]{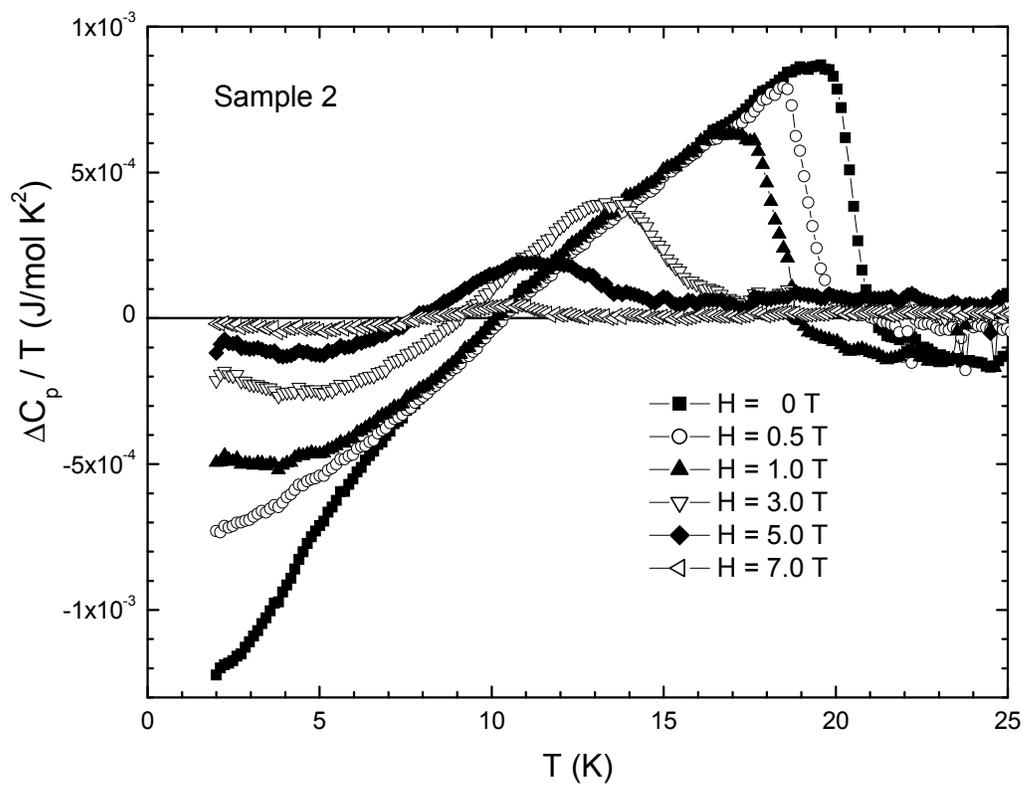}
\end{center}
\caption{Heat capacity difference $\Delta C_p/T =
(C_p(H)-C_p(9T))/T$ as a function of temperature for
Mg(B$_{0.8}$C$_{0.2}$)$_2$ --- 1100$^{\circ}$C/24h sample taken
for different applied fields.} \label{HC_2}
\end{figure}

One of the samples was chosen for more detailed measurements in
different applied magnetic fields. The results of these
measurements are presented in Fig. \ref{HC_2} in the form of
$\Delta C_p/T = (C_p(H)-C_p(9T))/T$ as a function of temperature.
The shift in the specific heat jump at superconducting transition
is a manifestation of the upper critical field and is consistent
with $H_{c2}(T)$ measured by other techniques (see stars in Fig.
\ref{Hc2}). The more interesting feature appears to be the low
temperature shoulder in the excess of specific heat seen in
$(C_p(H=0)-C_p(9T))/T$ data below 10 K (also clearly seen for both
samples in the previous figure). A similar feature was observed in
pure MgB$_2$ by different groups \cite{HC1,HC2,HC3} and was
interpreted as experimental evidence of a second, much lower
energy, superconducting gap in MgB$_2$ \cite{HC1,HC2,HC3,Choi}.
There are other important similarities between heat capacity data
of pure magnesium diboride and the carbon doped sample: the low
temperature feature disappears (lower gap is quenched) in small
(0.5 T) applied field and the $\Delta C_p/\gamma T_c$ value is
substantially smaller than expected for a BCS superconductors.
These two peculiar results were shown to be present in pure
MgB$_2$ and to be consequences of the two-gap nature of
superconductivity in this material. These similarities in heat the
capacity data of pure and carbon doped magnesium diboride imply
that despite the significantly suppressed $T_c$ and apparently
large increase in resistivity, the novel double gap nature of
supercondutivity persists in our nominal
Mg(B$_{0.8}$C$_{0.2}$)$_2$ samples. It is worth noting whereas
there have been estimates of $T_c \approx$ 20 K for "isotropic"
(single gap) MgB$_2$ \cite{liu,choiPRB} the case we seem to find
for our nominal Mg(B$_{0.8}$C$_{0.2}$)$_2$ appears to be quite
different with two distinct gaps. Further research will be
required to confirm this initial observation but, as it currently
stands, this finding requires that the two superconducting gaps
survive quite dramatic perturbations.

\begin{figure}
\begin{center}
\includegraphics[angle=0,width=140mm]{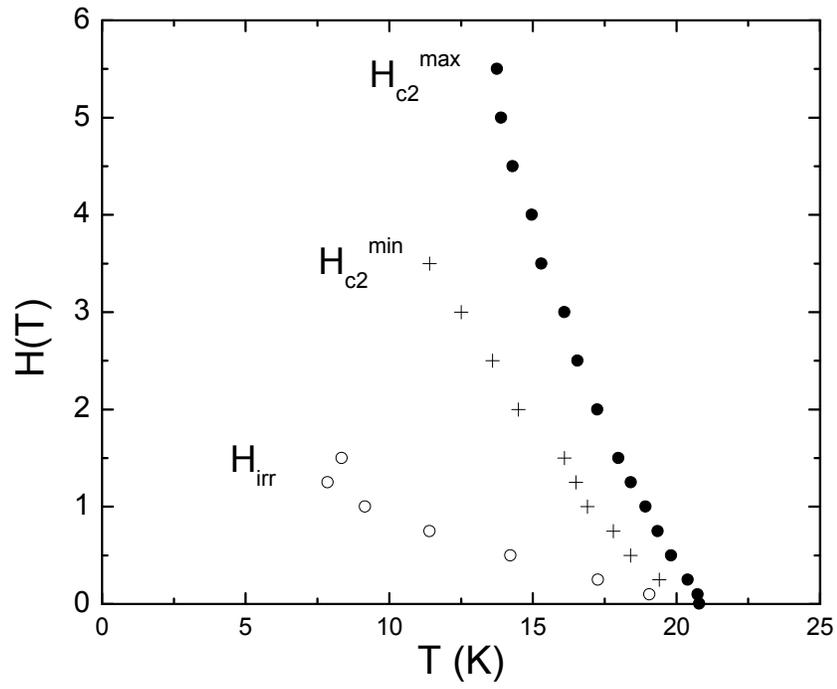}
\end{center}
\caption{Anisostropic $H_{c2}$ and irreversibility line $H_{irr}$
curves for Mg(B$_{0.8}$C$_{0.2}$)$_2$ --- 1100$^{\circ}$C/24h
sample.} \label{Ani}
\end{figure}

Finally, the anisotropic upper critical field for carbon doped
MgB$_2$ was evaluated from temperature dependent magnetization
measurements following the procedure outlined in
\cite{anisoPRB,anisoBRIF}. Although the feature corresponding to
$T_{c2}^{min}(H)$ in $(\partial M/\partial T)\mid_H$ for this
sample was slightly broader than for pure MgB$_2$ it was possible
to trace $H_{c2}^{min}$ above $\approx$ 12 K (see Fig. \ref{Ani}).
The anisotropy of $H_{c2}$ at $\approx \frac{2}{3} T_c$ is close
to 2, i.e. carbon doped MgB$_2$ has apparently less anisotropic
$H_{c2}$ than pure compound that may be a result of distortions in
the Fermi surface and require additional
theoretical/band-structure studies.

\section{Conclusions}

The synthesis of carbon doped magnesium diboride from magnesium
and boron carbide (B$_4$C) with a nominal stoichiometry of
Mg(B$_{0.8}$C$_{0.2}$)$_2$ was optimized and resulted in nearly
single phase material with $T_c \approx$ 22 K. Samples obtained by
this route have an upper critical field of $\approx$ 9 T at 10 K
and the slope of $H_{c2}(T)$ is much steeper than for pure
MgB$_2$. The sample has moderate $J_c$ values pointing out that
carbon introduced in the lattice via this synthetic route does not
increase pinning significantly. The specific heat data taken in
different applied fields suggest that the two gap
superconductivity is preserved in the Mg(B$_{0.8}$C$_{0.2}$)$_2$
sample despite the heavily suppressed $T_c$. In addition whereas
there is a significant $H_{c2}$ anisotropy ($\gamma$ $\approx$ 2
for $ T/T_c \approx \frac{2}{3}$), it is reduced from the
anisotropy found in pure MgB$_2$.

\section{Acknowledgements}

We would like to thank M. A. Avila and N.E. Anderson, Jr. for
helpful assistance and many fruitful discussions. Ames Laboratory
is operated for the U. S. Department of Energy by Iowa State
University under Contract No. W-7405-Eng.-82.  This work was
supported by the director for Energy Research, Office of Basic
Energy Sciences.

\newpage

\clearpage

\clearpage

\clearpage

\clearpage

\clearpage

\clearpage

\clearpage

\clearpage

\clearpage

\clearpage

\clearpage

\clearpage

\end{document}